# EFFECT OF SUB-STRUCTURE IN CLUSTERS ON THE LOCAL WEAK-SHEAR FIELD

PRIYAMVADA NATARAJAN
*Institute of Astronomy, University of Cambridge, Cambridge U.K.*

AND

JEAN-PAUL KNEIB
*Institute of Astronomy, University of Cambridge, Cambridge U.K.*

## 1. Abstract

Weak shear maps of the outer regions of clusters have been successfully used to map the distribution of mass at large radii. The effects of substructure in clusters on such reconstructions of the total mass have not been systematically studied. We propose a new method to study the effect of perturbers (bright cluster galaxies or sub-groups within the cluster) on the weak shear field. We present some analytic results below.

## 2. Analysis of the local weak shear field

Working in the coordinate frame of the perturber, we model the total cluster potential as the sum of a global smooth piece (assumed to be circular for illustration) and a perturbing piece which in general has both an elliptical part and a circular part.

$$\phi_{\mathbf{tot}} = \phi_{\mathbf{cluster}} + \phi_{\mathbf{perturber}}$$

$$\phi_{\mathbf{c}} = \phi_{\mathbf{oc}}\left(|\vec{r} + \vec{r_c}|\right)$$

$$\phi_{\mathbf{p}} = \phi_{\mathbf{op}}\mathbf{r}\left(1 + \frac{\epsilon}{6}\cos 2(\theta - \theta_0)\right)$$

where $\epsilon$ and $\theta_0$ are respectively the intrinsic ellipticity and orientation of the perturber. We calculate the shear in the weak limit at a given point as the sum of contributions from these two sources. In order to obtain a quantity that can be compared to observations we define an averaging procedure that involves integration over an annulus of finite radius which gives,

$$<g_x> = \int_0^{2\pi} \int_{r_1}^{r_2} g_x(r,\theta)\, r\, dr\, d\theta$$

$$<g_y> = \int_0^{2\pi} \int_{r_1}^{r_2} g_y(r,\theta)\, r\, dr\, d\theta$$

$$<g_x>_c = -\frac{\phi_{oc}}{2r} \; ; \; <g_y>_c = 0$$

$$<g_x>_p = -\frac{\phi_{op}}{12(r_1+r_2)} \epsilon \cos 2\theta_0$$

$$<g_y>_p = -\frac{\phi_{op}}{12(r_1+r_2)} \epsilon \sin 2\theta_0$$

The aim is to extract the parameters of the perturber ($\phi_{op}$ and $\epsilon$) independently, so we convolve with an appropriate window function $\hat{W}(\theta)$ that maximizes the signal. An optimum choice is the following window,

$$\hat{W}_x(\theta) = \cos 2\theta \; ; \; \hat{W}_y(\theta) = \sin 2\theta$$

Some interesting features which are primarily due to the particular choice of averaging procedure are that for a circular perturber and an elliptical perturber oriented at $\frac{\pi}{4}$ with respect to the cluster center the contribution to the local weak shear field as defined above vanishes.

## 3. Conclusions

Applications of this formalism provide us with a probe of the structure of cluster galaxies. With high resolution wide field data we can put limits on halo sizes and masses of cluster galaxies, which are relevant for understanding the details of the process by which clusters assemble. We shall also be able to quantify the errors in mass estimates from lensing due to the presence of substructure in clusters.